# On the Sustainability of Electrical Vehicles

Tai-Ran Hsu, ASME Fellow
Professor and Chair
Department of Mechanical Engineering
San Jose State University
San Jose, CA 95192-0087

ABSTRACT

Many perceive electric vehicles (EVs) to be eco-environmentally sustainable because they are free of emissions of toxic and greenhouse gases to the environment. However, few have questioned the sustainability of the electric power required to drive these vehicles. This paper presents an in-depth study that indicates that massive infusion of EVs to our society in a short time span will likely create a colossal demand for additional electric power generation much beyond what the US electric power generating industry can provide with its current generating capacity. Additionally, such demand would result in much adverse environmental consequences if the current technology of electric power generation by predominant fossil fuels continues. Other rarely accounted facts on environmental impacts by EVs are the substantial electric energy required to produce batteries that drive EVs, and the negative consequences relating to the recycling of spent batteries.

.

## I. INTRODUCTION

There is an unprecedented surge of interest in and the desire of producing gas-electric hybrid (HEVs) and all-electric vehicles (EVs) by the governments and public in the industrialized countries. This surge of interest and desire is attributed to several factors relating to the life-threatening air pollution in many parts of the world, the high petroleum price, conflicts in oil producing Middle East regions and the alarming trend of global warming due to rapid increase of greenhouse gas emissions by all sources, including those from transportation.

President Barack Obama shared his vision on HEVs and EVs in his statements made at a visit to Edison's "Garage of the Future" in Southern California on March 19, 2009; he would like to see a million plug-in HEVs (PHEVs) running at 150 miles per gallon of gasoline by Year 2015. Additionally, he would like to convert half the federal cars and trucks into either PHEVs or EVs by 2012. He had already pledged $12 billion in funding for research and development in EVs and additional $2.4 billion in batteries. The federal government had also offered a $7,500 tax credit to anyone purchasing a new EV. All the financial supports and the incentives for owning EVs are based on the perception that EVs are eco-environmentally sustainable means of transportation of the future.

The concept of design and construction of electricity-powered automobiles is by no means new. Active research and development in EVs to replace gasoline-powered vehicles can be traced back to the early 1970s amidst the first, but short-lived energy crisis in this country triggered by unilateral control of oil production and price set by the OPEC cartel (OPEC = Organization of the Petroleum Exporting Countries). There was a strong desire to produce non-gasoline powered vehicles by the automotive industry in this country, and EVs are among the top choice by this industry. Unfortunately, this desire diminished soon after consumers' acceptance of high petroleum price as a "fact-of-life." The next wave of strong interest in producing EVs did not occur until late 1980s resulted from the worsened air pollution in many urban centers in the world. The term "sustainable development" initiated by the United Nations [1] also renewed strong interest by citizens of industrialized countries in the increasing use of renewable energy sources for sustainable economic development. Because transportation consumes about 28% of total energy consumed in US and other industrialized nations [2], and it is also a major contributor to air pollution in major urban centers, EVs that are powered by "clean" electricity are perceived to be a viable sustainable means of transportation by industrialized countries. Consequently, the giant U. S. auto-maker, the General Motor Corporation pioneered in the design and construction of a battery-power vehicle called EV1 in 1996 in responding the renewed public interest. Unfortunately the EV1 vehicle never actually caught on by the consumers. It was subsequently removed from the marketplace three years later. There were a number of reasons for the failure of this venture, as described in a well-publicized documentary movie on "Who Killed the Electric Car" [3]. Fortunately, the strong desire in eco-environmental sustainability, and the pressing needs to mitigate greenhouse gas emissions by automobiles have sustained the continuous effort in developing EVs that would be economically viable, as well as marketable enough to replace most or all gasoline-powered vehicles in the new millennium.

This paper will address a serious but rarely publicized issue on whether EVs are indeed eco-environmentally sustainable means of transportation for the US with the prevailing technologies used in generating electric power that is required to drive these vehicles. There is no qualm about EVs being emissions free as far as eco-environmental sustainability is concerned. However, the electricity that is required to produce and charge the batteries for the EVs is generated by predominantly fossil fuels that emit not only toxic solid and gaseous byproducts to the atmosphere, but also the greenhouse gases for global warming. A conservative estimate based on slight outdated available information on the required additional electric energy to substitute all gasoline-powered vehicles with EVs is staggering; it indicates an equivalent additional electric energy required to produce the batteries and drive all household gasoline-powered vehicles in the US in 2001 would exceed the total electric power generating capacity by all US utilities combined in the entire Year of 2005. The potential adverse effects to eco-environment by massive infusion of EVs in the U.S. with prevalent electricity generation technologies are beyond anyone's imagination.

One may thus perceive that the sustainability of EVs is inseparable to the sustainability of electric power generation required to power these vehicles. EVs can be viewed as sustainable means of transportation only when the electricity that is required to drive them can be generated by clean renewable energies such as hydroelectric power, hydrogen fuel cells, hybrid solar photovoltaic and wind power, and ultimately nuclear fusion. The issue on effective recycling of millions spent batteries also needs to be dealt with if EVs will indeed be eco-environmentally sustainable.

## II. A GLANCE AT SUSTAINABLE DEVELOPMENT

The term "Sustainable Development" is a universally accepted solution to economic development of humankind for the present and future generations. It is a socio-economical process characterized by the fulfillment of human needs while maintaining the quality of the natural environment indefinitely. The concept of sustainable development came to general public awareness following the publication of a 1987 report by the Brundland Commission established by the UN General Assembly in 1983 [2].

Figure 1 illustrates the three principal pillars of sustainable development: "Social justice," "Environmental protection," and "Economic development." The pillars on "Environmental protection" and "Economic development" appear self-explanatory. The pillar on "Social justice," however, advocates the health and sustainability of social structure resulted from economic growth. These three pillars are equally important in implementing sustainable development as they are interrelated by "equitability," "bearability," and "viability" as indicated in Figure 1.

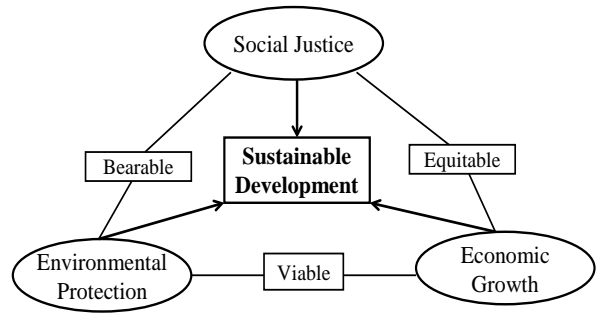

Figure 1 Three Pillars of Sustainable Development

The implementation of sustainable development requires the consumption of renewable resources provided by nature be kept less than what it can provide. Renewable resources include energy, quality air and water, as well as productive soil, ecoforestry, marine lives and all biospecies, as indicated in Figure 2.

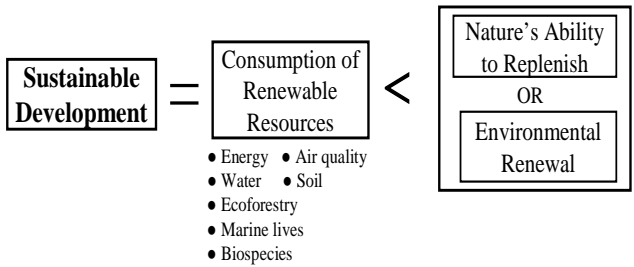

Figure 2 Essence of Sustainable Development

We realize the fact that "energy" and "quality air" are listed as the top two viable renewable resources in Figure 2. Many believe that EVs are typically sustainable because they are driven by electric power, which produces no emissions to pollute the air. They are thus ideal substitutes to the existing gasoline–powered vehicles, which not only pollute the air, but they also consume precious non-renewable petroleum resource. Unfortunately, the electricity that is required to produce and charge the batteries that drive the EVs and the PHEVs is currently generated by burning dirty non-renewable energy sources such as coal, oil and natural gas. Electric power generation is a major contributor to the worsening air quality in the world, and the producer of greenhouse gases for global warming. Electric vehicles (EVs) and PHEVs that are powered by electricity cannot be viewed as sustainable means of transportation with current technologies of electric power generations by burning fossil fuels.

## III. ELECTRIC POWER GENERATION IN USA

Of the total electric power generation at 13,650 Tera Watts-hour in the world in Year 2005 [1] (or 20,261 TWh in 2008) (1 Tera Watt (TWh) = $10^{12}$ Watts), 4055 billion kWh, or 4055 TWh were produced in the US (or 4369 TWh in 2008). Figure 3 shows the mix of energy sources that produced this amount of the electricity in the US [2]. One will readily observe that 71.4% of the electricity production in the US in 2005, or 2895 TWh was produced by burning fossil fuels of coal, oil and natural gas. Another 19.3% of the electricity, or 783 TWh was generated by non-polluting nuclear energy, but with serious environmental consequences by its spent fuels. One may thus conclude that over 90% of electricity generated in the US in 2005 was actually produced by non-renewable and environmentally unfriendly energy sources.

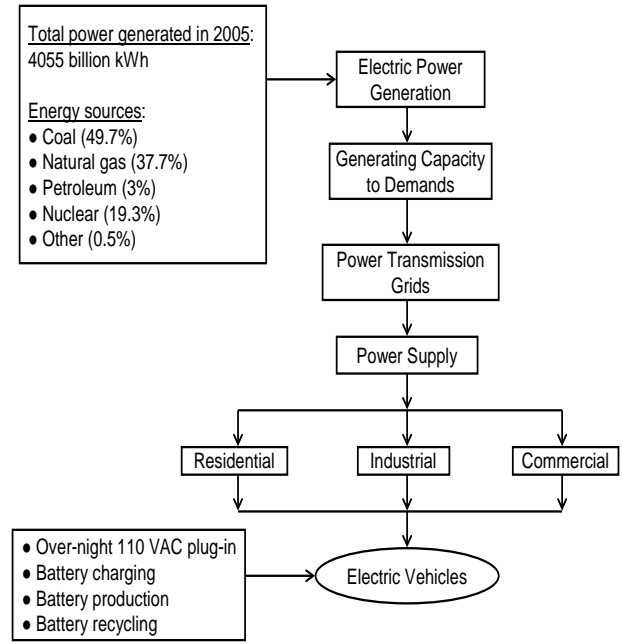

Figure 3 US Electric Power Generations by Mixed Energy Sources in 2005

## IV. ENERGY CONSUMPTION BY GASOLINE-POWERED VEHICLES

The total energy consumption in the US in 2005 was 29,000 TWh [2] (or 27,864 TWh in 2012). Of this amount of energy consumption in 2005, 28% was consumed by the transportation sector. It also indicated that 61% of that amount was by gasoline fuel. Thus, we may estimate the approximate energy consumed by gasoline-powered vehicles to be 29000 TWh x 0.28 x 0.61 = 4953 TWh – a staggering amount! If we view this amount of energy to be what was consumed by automobiles in that year, we would come up with a hypothesis on how much equivalent electric power would be required to convert all these gasoline-powered vehicles into EVs. This hypothesis obviously is not realistic because the total electric power generation in the US in that year was only 4055 TWh. Conversion of all gasoline-powered vehicles to EVs in that year would require US utilities to generate equal or more than double their generating capacity. Another set of statistical data [4] indicates that the household vehicles in the US consumed $113.1 \times 10^9$ gallons of gasoline in 2001. By using a conversion factor of 114000 Btu/US gallon of gasoline and 1 Btu = 0.2929 Watt-hour (Wh), we would reach an amount of 3778 TWh of energy consumed by household cars alone in this country in 2001. This slightly out-dated data on energy consumption by household cars has also shown the need for substantial additional electric power generation by US utilities if all these cars were to be converted into EVs in that year.

## V. ENERGY REQUIRED IN PRODUCTION OF BATTERIES FOR EVS

Batteries are essential parts for any electric vehicle (EV) or plug-in hybrid gas-electric vehicle (PHEV). They amount to 20 to 40% of the total mass of typical EVs [5]. So they consume significant amount of driving power of the vehicles. Battery innovations are the key to making PHEVs more eco-environmentally sustainable because lighter and easier-to-recharge batteries will improve efficiencies of these vehicles. They could also spark mass-produced PHEVs and even resurrect the idea of all-electric vehicles.

There are three issues related to the batteries in the sustainability of EVs:

1) The energy required to produce batteries,
2) Emissions in producing batteries, and
3) Energy and environmental consequences in recycling consumed batteries.

Table 1 presents properties of two of the four common batteries used in EVs presented in Reference [5]. A 25-kWh energy output is assumed for these batteries used for small EVs.

The authors of Table 1 indicated that average energy contained in each battery is about 25 kWh, and the energy required to produce these batteries are: 3430 kWh for each lead-acid battery, and 7176 kWh for each NiMH battery.

Table 2 lists the performance of ten emerging EVs [6].

Table 1 Comparison of Batteries for Common EVs

| Item | Lead-acid (Pb-Acid) | Nickel-metal hydride (NiMH) |
|---|---|---|
| Electrode materials | Lead on fiberglass mesh | Nickel hydroxide and metal hydride |
| Electrolyte | Sulfuric acid | Potassium hydroxide |
| Energy density (Wh/kg) | 50 | 75 |
| Mass for 25 kWh (kg) | 500 | 330 |
| *Energy to make (kWh)* | *3430* | *7176* |
| Significant emissions | Lead particulates | Unknown |
| Comments | Short battery life, existing recycling infrastructure | MH recycling process unknown |

Table 2 Performance of Ten Emerging EVs [6]

| Name of EV | Power, kW | Max. Speed mph | Max. Range, miles |
|---|---|---|---|
| Chevrolet Volt | 112 | 100 | 40 |
| Fisker Karmas | 300 | 125 | 50 |
| GM Opel Ampera | 112 | 100 | 37 |
| Mini E | 150 | 95 | 100-120 |
| Mitsubishi | 47 | 80 | 100 |
| Nissan E Car | 80 | | 100 |
| Tesla Roadster* | 215 | 125 | 227 |
| Th!nk city | 30 | 65 | 112 |
| Toyota Prius PHEV | Not available | Not available | Not available |
| ZENN | 22.4 | 25 | 30-50 |
| Mean | 118.7 | 90 | 91 |
| Median | 112 | 97.5 | 100 |

(http://www.forbes.com/2009/03/18/electric-car-new-lifestyle-vehicles-electric-cars.html)
*Powered by 6831 lithium-ion cells

We will use the "median values" in the listed performance of typical EVs in Table 2 for our subsequent analysis, as indicated in the last row in Table 2. From which, we have reached an estimate of (112 kW) x (100 mi)/(97.5 mi/h) ≈ 115 kWh electric energy required for each EV to run at 97.5 mph for 100 miles.

There are two ways one may estimate the energy required to produce the two kinds of batteries listed in Table 1 based on the estimated 4953 TWh energy required to replace all gasoline-powered vehicles derived from available data for 2005, or the 3778 TWh equivalent electric energy consumed by gasoline-powered household cars in 2001, as presented in Section IV.

1) Method A:

By using the estimated 115 kWh energy for each of the emerging EVs as calculated above, we will come up with an equivalent number of $43.07 \times 10^9$ EVs based on the 2005 statistics, or $32.85 \times 10^9$ EVs based on the 2001 statistics. On an assumption that each EV carries 4 batteries and we would come up with a total $172.28 \times 10^9$ batteries by using the 2005 statistics, and $131.4 \times 10^9$ batteries using the 2001 data.

2) Method B:

We may also estimate the energy required to produce batteries for EVs replacing the gasoline-powered vehicles in the following ways; Since each battery on average contains 25 kWh as mentioned above, we may estimate the total number of batteries to produce the energy required to replace all gasoline-powered vehicles in 2005 by electric power supplied by batteries as: $4953 \times 10^9 / 25 = 198.12 \times 10^9$ batteries based on statistics available for year 2005, or $3778 \times 10^9 / 25 = 151.12 \times 10^9$ batteries in Year 2001.

The required energy for producing batteries can be computed from the data shown in Table 3 using the estimated number of batteries required for replacing gasoline-powered vehicles in 2005 and 2001 respectively by EVs.

Table 3 Energy Required to Produce Batteries

| Batteries | Method A | | Method B | |
|---|---|---|---|---|
| | 2005 Data | 2001 Data | 2005 Data | 2001 Data |
| Lead-Acid Battery (Reqid energy:3430 kWh) | 591 TWh | 451 TWh | 679.55 TWh | 518.34 TWh |
| NiMH Battery (Reqid energy:7176 kWh) | 1236.45 TWh | 943.55 TWh | 1421.71 TWh | 1084.44 TWh |

One may readily observe from Table 3 that significant additional electricity energies up to 1421.17 TWh would be required to produce the batteries for EVs using NiMH batteries. This number is equivalent to 35% of the total 4055 TWh electric energy generated by the US utilities in 2005.

## VI. GREENHOUSE GAS EMISSIONS AND ENVIRONMENTAL CONSEQUENCES

According to Environmental Protection Agency of the US government, transportation sources accounted for approximately 29 percent of total U.S. greenhouse gas emissions (GHGs) in 2006. (http://www.epa.gov/otaq/climate/index.htm) It is the fastest growing source of greenhouse gas emission in the US. GHGs, accounting for 47 percent of the net increase in total U.S. emissions since 1990. Transportation is also the largest producer of the end-use source of $CO_2$, which is the most prevalent greenhouse gas. These estimates of transportation GHGs do not include emissions from additional lifecycle processes, such as the extraction and refining of fuel and the manufacture of vehicles, which are also a significant source of domestic and international GHGs productions.

In this section we will estimate how additional GHGs will result if we replace all gasoline-powered vehicles by EVs. Based on statistics indicated in a report [7], the total $CO_2$ emissions by electric power generation in US in 2005 was 2480 million metric tons with a total power generation at 4055 TWh [1]. In Sections IV and V, we have estimated that the energies required to replace all gasoline-powered vehicles by EVs would be the sum of 4953 TWh for the substitute power for the EVs with an additional 1421.17 TWh for the production of batteries. The total required additional electricity generation would thus be 6374.17 TWh, which would lead to an additional 3900 million metric tons of $CO_2$ emission based on the 2005 data. These additional $CO_2$ emissions would exceed half of the current total $CO_2$ emissions by this country.

Another major environmental impact with mass influx of EVs to the consumers market is the recycling of spent batteries. Authors of Reference [5] indicated that not all parts of batteries used in automobiles can be recycled. For example, the electrolyte of most batteries cannot be recycled and reused. In the case of NiMH batteries, the Ni electrodes can be recycled, but requires significant amount energy for recycling. Technologies for recycling the chunk metal hydrides remain unavailable. The environmental consequences to treat those estimated 40 billion used batteries will definitely be a major threat to the eco-environmental well being of the planet.

The production of batteries also produces significant emissions to the environment; in addition to the emission of toxic gases such as $SO_2$, there are other particulates including iron oxides, sulfur oxides, carbonaceous compounds and chlorides emitted to the air.

The astounding demand for additional electric power in mass influx of EVs or PHEVs to our society will inevitably pose serious demand for other precious natural resources such as fresh water. Sources indicated that coal-fire and natural gas-fired power plants spend on average 480 and 180 gallons of fresh water respectively for every MWh of electric energy that the utilities produce. An estimated 4953 TWh additional electric energy for replacing all gasoline-powered vehicles in 2005 with the fuel sources for electricity generations in that year as shown in Figure 3, would come up with 1181.58 trillion gallons of fresh water consumption by coal fired stations and 336.11 trillion gallons of fresh water used in electricity generation by natural gas-fired stations in that year. Coal-firing electric power generation also contains toxic waste of mercury and arsenic, which post major threat to human health and marine lives.

## VII. STRATEGY FOR SUSTAINABLE ELECTRIC VEHICLES

The statistic data presented in this paper though are slightly outdated but they have adequately demonstrated that the issue on the sustainability of EVs (and PHEVs) cannot be dissociated from the sustainability of electric power generations by utilities. While much effort have already been initiated to generate electricity by clean coal technology [8] with smart grid distribution systems for efficient power transmission and distributions [9], a great deal more effort is required to produce electricity by clean renewable energy sources such as solar and wind. A new US government initiative on "20% wind energy by 2030" [10] targets having 20% of the electric power generated by this country is from wind energy. The same goal was set by the US-based photovoltaic idustry [11]. Solar and wind are only two known clean eco-environmentally sound energy sources for electric power generation. Current national goals for having electricity generation by both these clean renewable energy sources thus appears to be 30% of the toal electricity power generation by these sources by Year 2030. If one uses the 4055 TWh electric energy generation by the US utility in 2005 as a base line capacity, we would expect 30% of that amount, i.e. 1216 TWh by solar/wind power combined by Year 2030. This would mean 1216/4953 = 0.25 or 25% of all household gasoline-powered vehicles in that year could be covnerted to EVs without producing additional environment-threatening emissions, and thus sustainable vehicle conversion. This rate of gradual introduction of EVs and PHEVs to the US market would allow the utilities in this country with the necessary time to supply additional electric power generation by clean renewable energy sources required to drive EVs and PHEVs for the US consumers.

## VIII. SUMMARY AND CONCLUDING REMARKS

A rather conservative estimate based on slightly outdated available information on the required additional electric energy for replacing all gasoline-powered vehicles to EVs is staggering; it indicates an equivalent additional electric energy of 3778 TWh required to replace all household gasoline-powered vehicles in the US in 2001 with EVs, with additional 1236 TWh electric energy required to produce the batteries for these EVs. The total additional electric energy required for such replacement would exceed the total electric power generating capacity of 4055 TWh by all US utilities combined in the entire Year of 2005.

The consequence of the staggering additional electric energy required to power large scale new EVs as presented above is just one factor, other issues relating to haste introduction of EVs are the over-loading the vital grid transmission of electric power as well as the patterns of peak-power generations that have been established and practiced by U.S. utilities for many years.

The analyses presented in this paper though are based on slightly outdated statistical data on electric power generations and the consumption of energy by gasoline powered vehicles and electric battery productions they nevertheless have indicated that any ill planned and unregulated massive introduction of electric vehicles without thoroughly planned electric power generations and distributions by clean renewable energy would result in not only eco-environmental disaster but also in serious socio-economic problems because of insufficient power supply to the demand by the nation's industry and business. Government, utilities and EV industry need to work together to develop a credible road map on the nation's striving towards sustainable transportation, which is vital to its citizen's livelihood and also to the eco-environment sustainability of the global village.

BIOGRAPHY

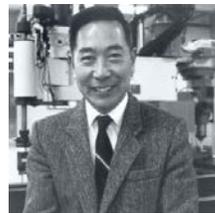

Tai-Ran Hsu, a Fellow of ASME, is currently a Professor and the Chair of the Department of Mechanical Engineering at San Jose State University. He received a BS degree from National Cheng-Kung University in Taiwan, China; MS and Ph.D. degrees from University of New Brunswick and McGill University in Canada respectively. All his degrees were in mechanical engineering. He worked for steam power plant equipment and nuclear industries prior joining the academe. He taught mechanical engineering courses and served as department heads at universities in both Canada and USA. He has published over 120 technical papers in peer reviewed systems and eight books in finite element method in thermomechanics, computer-aided design, and a well-received textbook on microelectromechanical systems (MEMS) design and manufacture. The second edition of the latter book was published in March 2008.